\documentclass[lettersize,journal]{IEEEtran}
\usepackage{cite}
\usepackage{amsthm}

\usepackage{multirow}
\usepackage{braket}
\usepackage{lineno,xcolor}
\usepackage{amsmath}
\usepackage{amsfonts}
\usepackage{amssymb,scalerel,verbatimbox}
\usepackage{wasysym}
\usepackage{stmaryrd}
\usepackage{bbm} 
\usepackage{tabularx}
\usepackage{multirow}
\usepackage{mathtools}
\usepackage{enumerate}
\usepackage{subfigure}

\usepackage{lipsum}
\usepackage{cuted}
\usepackage{flushend}
\usepackage{pgf}
\usepackage{bm}
\usepackage{multicol}
\usepackage{color}
\usepackage{url}
\usepackage{tabularx}
\usepackage{bibunits}
\usepackage{wrapfig}
\usepackage{sidecap}
\usepackage{soul,xcolor}
\usepackage{eurosym}
\usepackage{mathrsfs}
\usepackage[utf8]{inputenc}
\usepackage{algorithmicx}
\usepackage{algpseudocode}
\usepackage{algorithm}
\usepackage{stmaryrd}
\usepackage{upgreek}
\usepackage[hidelinks]{hyperref}
\usepackage{paralist}
\usepackage{graphicx}        
\usepackage{subcaption}  
\usepackage{setspace}
\usepackage{tikz}
\usetikzlibrary{arrows}
\usepackage{cleveref}
\usepackage{etoolbox}
\makeatletter
\patchcmd{\@makecaption}
{\scshape}
{}
{}
{}
\makeatother

\DeclareMathAlphabet{\mathpzc}{OT1}{pzc}{m}{it}

\usepackage{epstopdf}
\usepackage{soul}
\setstcolor{red} 
\setul{0pt}{0.7pt}

\newtheorem{defn} {Definition}

\IEEEoverridecommandlockouts

\DeclareMathOperator*{\argmin}{arg\,min}

\usepackage{enumitem}
\setlist{nolistsep}
\newcommand*\xor{\oplus}

\usepackage[left=0.59in,right=0.59in,top=0.6in,bottom=0.6in]{geometry}
\usepackage{bbm}
\begin{document}
 \title{Collective Bit Flipping-Based \\Decoding of Quantum LDPC Codes}
\author{Dimitris~Chytas,  Nithin~Raveendran,~\IEEEmembership{Member,~IEEE,}  and Bane~Vasi\'{c},~\IEEEmembership{Fellow,~IEEE}
}

\maketitle
\begin{abstract}
Quantum low-density parity-check (QLDPC) codes have been proven to achieve higher minimum distances at higher code rates than surface codes. However, this family of codes imposes stringent latency requirements and poor performance under iterative decoding, especially when the variable degree is low.
In this work, we improve both the error correction performance and decoding latency of variable degree-$3$ ($d_v$-$3$) QLDPC codes under iterative decoding.
Firstly, we perform a detailed analysis of the structure of a well-known family of QLDPC codes, i.e., hypergraph product-based codes. Then, we propose a decoding approach that stems from the knowledge of harmful configurations apparent in these codes. Our decoding scheme is based on applying a modified version of bit flipping (BF) decoding, namely two-bit bit flipping (TBF) decoding, which adds more degrees of freedom to BF decoding. The granularity offered by TBF decoding helps us design sets of decoders that operate in parallel and can collectively decode error patterns appearing in harmful configurations of the code, thus addressing both the latency and performance requirements. Finally, simulation results demonstrate that the proposed decoding scheme surpasses other iterative decoding approaches for various $d_v$-$3$ QLDPC codes.
\end{abstract}	

\begin{IEEEkeywords}
QLDPC codes, bit flipping decoding, hypergraph-product codes, symmetric stabilizers, trapping sets, error-floor.
\end{IEEEkeywords}

\section{Introduction}

\IEEEPARstart{
Q}{uantum} low-density parity-check (QLDPC) codes have gained attention recently over topological codes because of their superiority in the minimum distance and their code rate scaling asymptotically~\cite{panteleev2021quantumLinearMinDLocalTestable, QuantumTannerCodeszemor}. However, QLDPC codes impose long-range qubit connections, which can increase noise and also induce additional delays\cite{xu2024constant, bravyi2024high},  
potentially further restricting the latency budget available for decoding implementations. These latency requirements are also accompanied by poor error correction performance, which is apparent for 
iterative decoders such as belief propagation (BP) or min-sum (MS). Failure of iterative decoding is mainly attributed to inherent properties of QLDPC codes, such as degeneracy and the presence of short cycles 
~\cite{Poulin_2008,hanzo_15_years}. This performance degradation, especially in the error-floor region, is mainly observed in the case of low variable-degree QLDPC codes, where iterative decoding fails to correct low-weight error patterns that appear inside harmful structures of the code, referred to as trapping sets~\cite{quantumTS}. 

Several approaches in the literature modify the message-passing rules or message scheduling to break symmetries imposed by symmetric stabilizers and improve performance. 
For example, Poulin and Chung~\cite{Poulin_2008} proposed heuristic modifications to BP, called freezing, random perturbation, and collision, which improved the performance of highly degenerate codes.
However, these approaches are not deterministic and do not fully utilize the code structure. 
Recently developed BP with bias using Oscillating Trapping Sets (BP-OTS)~\cite{BPOTS} improves performance after modifying the decoding rules for variable nodes that participate in trapping sets of topological codes. However, it has not been demonstrated for other families of QLDPC codes. 
Regarding the scheduling of iterative decoders, various approaches that take into account the structure of QLDPC codes have been proposed. Layered BP and MS algorithms~\cite{layeredQuantum} improve message-passing decoding performance because they are able to exploit the degeneracy. However, their serial decoding nature imposes additional latency requirements. 
Similarly, advanced post-processing techniques such as ordered statistics decoding (OSD)~\cite{osd}, stabilizer inactivation (SI)~\cite{StabInactivation_Julien_2022} decoding, and BP with guided decimation~\cite{decim} can improve decoding performance by paying the price of latency and high complexity, i.e., $\mathcal{O}(n^3)$, $\mathcal{O}(n^2log{n})$, and  $\mathcal{O}(n^2)$ respectively, where $n$ is the blocklength of the code. 
Finally, decoding approaches such as refined/memory BP decoding~\cite{refinedBP_QLDPC_2020, memory} take into account the correlation of $X$ and $Z$ errors, but significant performance results are obtained only when adaptive decoding approaches for choosing normalization parameters are combined with serial scheduling.

BF decoding, introduced in~\cite{63Gallager}, is a low-complexity alternative compared to other iterative algorithms and is widely used in classical LDPC decoding, where it provides good trade-offs between performance and latency-complexity~\cite{compBF}. 
These properties make BF particularly suited for quantum error correction (QEC), and different variants of BF decoders are proposed. 
Notably, the small-set-flip (SSF) decoder~\cite{quantum_expander_codes} deals with degeneracy (see definition of degeneracy in Section~\ref{sec:tbf}) by \emph{exhaustively flipping sets of bits contained in the support of stabilizers}, instead of single bits. This approach introduces asymmetry, which helps decode degenerate error patterns. 
However, SSF provides guaranteed error correction for codes constructed by the hypergraph product of expander codes. Therefore, very large codes are required for SSF to be effective in practical scenarios.
Moreover, SSF is not guaranteed to succeed in decoding error patterns appearing inside harmful structures except symmetric stabilizers. 
SSF decoder has also been used as a successor to BP (post-processing technique)~\cite{grospellier2021combining}, where BP is deployed in order to reduce the weight of the error pattern before it is decoded by SSF. Recently, targeting low-latency decoding, the Turbo-XZ algorithm deploys BF decoding with iteration-dependent thresholds in order to correct error patterns appearing in certain trapping sets~\cite{turbo}. However, this approach cannot take advantage of the degeneracy of the code. BF techniques have also been used in the scope of topological codes. 
Progressive-proximity bit flipping~\cite{pacenti2024progressiveproximity} is able to decode topological codes by exploiting certain structural properties. 

Our proposed decoding scheme is also based on BF, but differentiates from the previous approaches in the sense that it specifically targets low-latency decoding combined with guaranteed correction of error patterns appearing inside specific structures of QLDPC codes.
More specifically, we deploy \emph{two-bit bit flipping} (TBF) decoding, which has been explored before in the classical literature~\cite{TBFA}, and it has been proven to obtain superior performance compared to the original BF decoding.
The key feature of the TBF decoding is that it assigns more states to both variable and check nodes; therefore, variable nodes are treated with higher granularity than conventional flipping decoders,  which, in turn, helps in improving the error correction capability. Moreover, unlike other BF alternatives like the gradient-descent bit flipping (GDBF)~\cite{GDBF_Wadayama}, its probabilistic versions, and GDBF with momentum~\cite{gdbfMom}, the TBF decoder can be implemented by low-complexity deterministic functions that facilitate analysis and collective decoding. 
 
The concept of collective decoding has been explored in classical and quantum error correction literature. 
Regarding the former, a set of diverse low-complexity decoders operating in parallel\cite{FAIDdiv} achieves guaranteed error correction performance in a small number of iterations.
Relating to QEC, collective BP decoding was used within the framework of deep neural network-based learning \cite{pradhan2023learning} wherein a trapping set-based importance sampling approach is used to learn new decoding rules.

Our collective decoding approach is solely defined based on the structure of the code. 
We keep in mind that two major types of harmful configurations impede iterative decoding, namely, \emph{classical and quantum trapping sets} (or symmetric stabilizers), defined in Section.~\ref{sec:trapping}. 
Hence, the decoding framework should be able to deal with error patterns that appear in those two structures. 
Our approach is to first enumerate harmful configurations by using the shortest cycles of the code as a starting point and then exploit the parent-child relationships of these graphs in order to obtain the most dominant harmful graphs.
After the enumeration process, the goal is to develop a set of diverse TBF decoders capable of decoding all error patterns within quantum trapping sets and all error patterns up to a certain weight within classical trapping sets.

Our focus is on variable degree-$3$ ($d_v$-$3$) QLDPC codes from the generalized hypergraph product (GHP) code~\cite{osd} family. GHP codes suffer from error-floor under iterative decoding, which performs poorly despite the large distance of the codes. Nevertheless, this family of codes can achieve higher distance when compared to hypergraph product (HP) codes. Also, as was illustrated in~\cite{osd}, when GHP are decoded with BP-OSD, they can achieve better performance free of error-floor than HP codes.
This fact demonstrates the potential of this family of codes and enables us to analyze them. 
We also note that our restriction of the variable node degree is only for the sake of analysis, and we remark that the method is extendable to higher variable node degree codes.
Interestingly, we observed that for a family of $d_v$-$3$ QLDPC codes, the $6$-cycles lead to classical trapping sets  which are only interconnected by quantum trapping sets which in turn emerge from the $8$-cycles.
Such analysis provides the overview of the structure of $d_v$-$3$ QLDPC codes. Based on the knowledge of harmful structures in this family of codes, we provide a systematic method of generating the set of TBF decoders able to collectively correct low-weight error patterns appearing inside both classical and quantum trapping sets.

Finally, we provide simulation results featuring the selected set of TBF decoders, and we show that it can surpass the performance of message-passing decoders such as MS and layered MS for three different $d_v$-$3$ QLDPC codes. 

The main contributions of the paper can be summarized below:

\begin{enumerate}

    \item We analyze the structure of $d_v$-$3$ GHP codes, a family of codes with good error correction properties but poor performance under iterative decoding.
    \item The analysis reveals the interconnection of quantum and classical trapping sets and shows that these codes, and in general quantum codes, facilitate collective decoding.
    \item By using collective decoding, we are able to both correct error patterns that appear inside quantum trapping sets, thus exploiting degeneracy, and also correct those that appear in classical trapping sets.
    \item Performance results show improved performance over the MS decoding while also demonstrating the potential for further improvements because of the structure of the codes.
    \item To our knowledge, this marks the first attempt toward good performance/low latency under non-serial (or adaptive or post-processing) decoding techniques.
\end{enumerate}

The rest of the paper is organized as follows. In Section \ref{sec:pre}, we introduce the preliminaries of QEC along with an overview of syndrome BF decoding and classical and quantum trapping sets. In Section \ref{sec:structure}, we analyze the structure of $d_v$-$3$ GHP codes. In Section \ref{sec:tbf}, we present syndrome-based TBF decoding, which is extended in Section \ref{sec:dec}, where we introduce TBF-based collective decoding, and we propose a set of decoders that can correct up to weight-$5$ error patterns that appear inside certain classical trapping sets and all error patterns that appear inside quantum trapping sets. Finally, Section \ref{sec:sim} presents simulation results obtained over various $d_v$-$3$ QLDPC codes.

\section{Preliminaries}
\label{sec:pre}
\subsection{Stabilizer Formalism}
Consider the $n$-fold Pauli group,
\[
\mathcal{G}_n  \triangleq \{pP_1 \otimes \cdots \otimes pP_n: p \in \{\pm 1, \pm i \}, P_j \in \{I, X, Y, Z\} \},
\] where $I$, $X$, $Y$, and $Z$ are called \textit{Pauli operators}. 
Every element in $\mathcal{G}_n$ has eigenvalues $\pm 1$, and any two elements in $\mathcal{G}_n$ either commute or anticommute with each other. 
A stabilizer group $\mathcal{S}$ is an Abelian subgroup in $\mathcal{G}_n$. 
If $\mathcal{S}$ is generated by $n-k$ independent generators, it defines a $\llbracket n,k,d_{\text{min}} \rrbracket$ stabilizer code $C$ that encodes $k$ logical qubits into $n$ physical qubits, with $d_{\text{min}}$ being its minimum distance.
The elements of $\mathcal{S}$ are called stabilizers.
The set of generators of $\mathcal{S}$ can be represented by the stabilizer matrix $H$, whose $(i,j)^{\text{th}}$ element is given by the Pauli operator corresponding to the $j^{\text{th}}$ qubit in the $i^{\text{th}}$ stabilizer. For a more thorough description of the stabilizer formalism, one can refer to~\cite{Gottesman97}.

By applying the binary mapping to the stabilizer matrix, $S$, we obtain a $(n-k) \times 2n$ binary matrix:
\begin{equation}
    H = [H_X\ |\ H_Z],
\end{equation}
which we call the \textit{parity check matrix} of $C$. 
Since the corresponding stabilizers of $S$ commute with each other, it can be verified that~\cite{Gottesman97}:
\begin{equation}
\label{eq:commutativity}
    H_XH_Z^T + H_ZH_X^T = \mathbf{0}.
\end{equation}

In this work, we address HP codes, which constitute a class of Calderbank-Shor-Steane (CSS) codes. A $\llbracket n, k_X - k_Z, d_{\text{min}} \rrbracket$  CSS code~\cite{calderbank1996quantum_exists} is a stabilizer code constructed using two classical codes, $C_X [ n, k_X , d_{\text{min},X} ]$ and $C_Z [n, k_Z, d_{\text{min},Z}]$, where $d_{\text{min}} \geq \text{min}\{d_{\text{min},X} , d_{\text{min},Z}\}$~\cite{hypergraphProductCodeTillich} and $C_Z \subseteq C_X$. Given the stabilizer matrix $H_X$ of $C_X$ and $H_Z$ of $C_Z^{\bot}$ , the CSS code constructed from $C_X$ and $C_Z$ has the form $$H=\begin{bmatrix}
    H_Z & \mathbf{0}\\
    \mathbf{0} & H_X
\end{bmatrix},$$ where $H_X H_Z^T=\mathbf{0}$ (unless specified otherwise, we assume all operations on binary matrices and vectors are performed on the binary field).

CSS codes facilitate binary decoding because of their structure, i.e., $X$ errors are decoded using $H_Z$, and $Z$ errors are decoded using $H_X$. Therefore, it is sufficient to consider the two independent binary symmetric channels (BSCs) rather than the
depolarizing channel, thus ignoring the correlation between the $X$ and $Z$ errors~\cite{mackay_quantum}. Let us denote as $\mathbf{e}=[\mathbf{e}_X, \mathbf{e}_Z]$ the binary representation
of a Pauli error acting on the $n$ qubits. The corresponding input syndromes are obtained as $\mathbf{s}_Z= \mathbf{e}_X  H_Z^T$
and $\mathbf{s}_X= \mathbf{e}_Z  H_X^T$, or in a compact form, $\mathbf{s}=[\mathbf{s}_X, \mathbf{s}_Z]$. 

An all-zero syndrome vector indicates that all stabilizers commute with the error pattern. On the other hand, when the syndrome vector is non-zero, an error has been detected. The goal of a syndrome-based decoder is to produce an estimate error pattern, $\mathbf{\hat{e}}$, which corresponds to a syndrome $\mathbf{\hat{s}}$ that matches $\mathbf{s}$. We observe a decoding success if the initial error $\mathbf{e}$ is recovered up to a stabilizer, which means that $\mathbf{e} 
 \xor \hat{\mathbf{e}}$ belongs to the rowspace of the parity check matrix $H$. 
Error correction is unsuccessful when the decoder is not able to match the input syndrome $\mathbf{s}$, or when the decoding process results in a \textit{logical error}, 
such that $\mathbf{e} \oplus \hat{\mathbf{e}}$  commutes with all the stabilizers, but it's not in the rowspace of $H$.

The $H_X$ and $H_Z$ stabilizer matrices can each be represented as a \textit{Tanner graph}, denoted as $\mathcal{G}$, which is a bipartite graph with two sets of nodes:  $n$ variable (qubit) nodes $\mathcal{V} = \{v_1,...,v_n\}$ and  $m$ check nodes $\mathcal{C} = \{c_1,...,c_{m}\}$. 
The set of edges (non-identity entries of the stabilizer matrix) connecting the two sets of nodes is denoted as  $\mathcal{E}$.
The degree of a node is the number of its neighbors. 
A $d_v$-variable-regular LDPC code has a Tanner graph in which all variable nodes have the same degree, $d_v$. 
The degree of a check node $c$ is denoted by $d_c$. 
For a Tanner graph $\mathcal{G}$, the girth $g$ is the length of the shortest cycle in $\mathcal{G}$.
For the rest of this paper, only one type of error ($X$ error) and its decoding will be considered; without loss of generality, the notation $H$ will refer to $H_Z$, $\mathbf{e}$ will refer to $\mathbf{e}_X$, and $\mathbf{s}$ will refer to $\mathbf{s}_Z$.

\subsection{Bit Flipping Decoding}
Before we delve into TBF decoding, we will describe the syndrome version of the well-known parallel BF algorithm, which is one of the simplest hard decision decoders for LDPC codes on the BSC~\cite{zyablov,96SS}.
The algorithm iterates until the input syndrome has been matched or until a maximum number of $L$ iterations is reached. We define $s_c$ to be the input syndrome value of the check node $c \in \mathcal{C}$.
The check node $c$ is said to be satisfied when the estimated syndrome value $\hat{s}_c$ is the same as the input syndrome value $s_c$. Otherwise, the check is said to be unsatisfied. 
This is modeled by the residual syndrome value $r_c$, which computes the XOR between the input and the estimated syndrome. 
In each iteration, the algorithm ``flips” all variable nodes that are connected to more unsatisfied than satisfied check nodes, i.e., flips $v$ if $\chi_1^{\ell}(v) > \frac{d_v}{2}$. 
Here, $\chi_0^{\ell}(v)$ and $\chi_1^{\ell}(v)$, respectively, denote the number of satisfied and unsatisfied check nodes that are connected to $v$ at the beginning of the $\ell^{\text{th}}$ iteration. 
In this paper, as we primarily focus on regular QLDPC codes, the degree $d_v$ is the same for every $v \in \mathcal{V}$. 
In order to adapt to the quantum decoding problem, we will consider a syndrome version of BF decoder, which iteratively flips variable nodes until it finds an error pattern $\hat{\mathbf{e}}$ with syndrome $\hat{s}=\hat{\mathbf{e}}H^T$ that matches with $s$, or until it reaches the maximum number of iterations $L$. 
The syndrome BF decoder is outlined in Algorithm~\ref{alg:bfa}. We denote as $\mathbf{s}=(s_1, s_2, \cdots, s_m)$ the input syndrome to the iterative decoder. The decoding output at the $\ell^{\text{th}}$ iteration is denoted as $\hat{\mathbf{e}}^{\ell}=(\hat{e}_1^{\ell}, \hat{e}_2^{\ell}, \cdots, \hat{e}_n^{\ell})$, $\ell \leq L$ in the following.  
\begin{algorithm}
    \caption{Syndrome-Based BF Algorithm}
     \label{alg:bfa}
  \textbf{Input}: $\mathbf{s}, H$, $L$ 

  \textbf{Output}: $\hat{\mathbf{e}}$
  \begin{algorithmic}[1]
      \State $\ell\gets 0$ \Comment{Initialization}
      \State $\hat{e}_v^{\ell} \gets 0$, $\forall v \in \mathcal{V}$ 
      \State $\mathbf{r}^{\ell} \gets \mathbf{s}$ \Comment{Residual syndrome}
      \While{$\mathbf{r}^{\ell} \neq 0$ and $\ell \leq L$}
     \State $\mathbf{\chi}_1^{\ell} \gets \mathbf{r}H$ \Comment{Number of unsatisfied checks}
      \For {$v \in \mathcal{V}$}
      
      \If {$\chi_1^{\ell}(v) > \frac{d_v}{2}$}
      \State $\hat{e}_v^{\ell+1} \gets \hat{e}_v^{\ell} \xor 1$ \Comment{Flip}
       \EndIf
       \EndFor
      \State $\hat{s}^{\ell+1}=\mathbf{\hat{e}}^{\ell+1}H^T$ \Comment{Estimated syndrome}
      \State  $r_c^{\ell+1}=s_c\xor\hat{s}_c^{\ell+1}$ , $\forall c \in \mathcal{C}$ 
       \State $\ell\gets\ell+1$
      \EndWhile
  \end{algorithmic}
\end{algorithm}

\subsection{Classical and Quantum Trapping Sets}
\label{sec:trapping}
In general, the failures of BF decoding can be characterized by the notion of trapping sets~\cite{5437420},
which will be used for the rest of the paper. We note that definitions given below are adopted by~\cite{ontology} and~\cite{quantumTS}. According to the quantum decoding problem, a variable node $v$ is said to be eventually converged 
if there exists a positive integer $I$ such that for all $\ell \geq I$, $\hat{e}_v^{\ell}=\hat{e}_v^{\ell-1}$. Also, a check node $c$ is eventually satisfied if there exists a positive integer $I$ such that for all $\ell \geq I$, $\hat{s}_c^{\ell}=s_c$.

\begin{defn}
For a syndrome-based decoder, a trapping set $\mathcal{T}$, is a non-empty set of variable nodes that are not eventually converged or are neighbors of the check nodes that are not eventually satisfied.
\end{defn}

If the sub-graph induced by the set of variable nodes $\mathcal{T}$ contains $a$ variable nodes and $b$ odd-degree check nodes, then  $\mathcal{T}$ is labeled as a $(a, b)$ trapping set. 
As in the Tanner graph representation, the variable nodes and check nodes of the trapping set-induced subgraph are represented by circular and square nodes, respectively. The odd-degree check nodes are shown using black-shaded square nodes.

A particular family of trapping sets, i.e., fixed sets, are detrimental to the BF decoding performance. 
Based on definition $4$ given in~\cite{ontology}, a trapping set is considered to be a fixed set or fixed point of the bit flipping algorithm, if the hard decision of the set of variable nodes $\mathcal{T}$ remains unchanged after one round of decoding.
Authors in~\cite{5437420} also observed that for $\mathcal{T}$ to be a fixed set, no variable node $v \in \mathcal{T}$ should be connected to more odd-degree than even-degree checks. 
If a fixed set containing $a$ variable nodes exists, then BF decoding is not guaranteed to correct all error patterns up to $a$ errors. 
Examples of fixed sets apparent in families of GHP codes are given in Fig.~\ref{fig:childrenof3}\subref{fig:33} and Fig.~\ref{fig:childrenof3}\subref{fig:86} respectively.

In addition to classical type trapping sets described above, QEC introduces another type of harmful configuration.
As we mentioned, in the quantum decoding problem, the decoder needs to converge to an estimate error pattern $\mathbf{\hat{e}}$  such that $\mathbf{\hat{e}} \xor \mathbf{e}$ is a stabilizer. This differs from the classical decoding problem where an exact match of error is required. 
We refer to error vectors $\mathbf{{e}}$ and $\mathbf{f}$ as \emph{degenerate} errors if $\mathbf{{e}} \xor \mathbf{f}$ is a stabilizer. Hence, it is equivalent to output any one of the degenerate errors as the candidate error pattern for matching the syndrome. 
However, in QLDPC codes whose minimum distance is higher than their stabilizer weight, certain degenerate errors can impede iterative decoding. 
A symmetric topology of the stabilizer sub-graph that contains degenerate error patterns $\mathbf{e}$ and $\mathbf{f}$ of equal weight will result in a decoding failure \cite{quantumTS}.
In the following sections, we will illustrate examples of such decoding failures when the iterative decoder attempts to converge to error patterns $\mathbf{e}$ and $\mathbf{f}$ simultaneously, thus not matching the input syndrome. 
This failure is caused by the symmetry of both the stabilizer and the iterative decoder update rules.
Hence, such errors are referred to as symmetric degenerate errors, and the corresponding sets of variable nodes as symmetric stabilizers or quantum trapping sets. 
Below, we give the definition of symmetric stabilizers adopted from~\cite{quantumTS}. 
\begin{defn}
A symmetric stabilizer is a stabilizer with a set of variable/qubit nodes whose induced sub-graph has no odd-degree check nodes, and that can be partitioned into an even number of disjoint subsets so that (a) sub-graphs induced by these subsets of variable nodes are isomorphic, and (b) each subset has the same set of odd-degree check node neighbors in its induced sub-graph.
\end{defn}

Assuming that a code has girth $g$, it is well known that the shortest cycles of the code, namely the ones with length $g$ and $g+2$, constitute trapping sets for iterative decoding algorithms~\cite{ontology}. 
Therefore, a way to obtain larger structures that are harmful to decoding is to exploit the parent-child relationship between trapping sets. 
We say that a trapping set $\mathcal{T}_1$ is a parent of $\mathcal{T}_2$ (or that $\mathcal{T}_2$ is a child of $\mathcal{T}_1$) if $\mathcal{T}_2$ contains a subset whose induced sub-graph is isomorphic to $\mathcal{T}_1$\cite{ontology}. 
We will use this idea in the following section to present the trapping set analysis of GHP codes.
We note here that for the rest of the paper, we will refer to symmetric stabilizers as quantum trapping sets interchangeably.

\section{Trapping Set Analysis of Generalized Hypergraph Product Codes}
\label{sec:structure}

HP codes is a family of QLDPC codes constructed by any two classical codes $C_{a_1} [n, k_{a_1} , d_{a_1}]$ and $C_{a_2} [n, k_{a_2} , d_{a_2}]$~\cite{hypergraphProductCodeTillich}. More specifically, supposing that $a_1$ and $a_2$ are the parity check matrices of the codes $C_{a_1}$ and $C_{a_2}$ (with dimensions $m_{a_1}\times n_{a_1}$ and $m_{a_2}\times n_{a_2}$ respectively), then the hypergraph product code is the code with $H_X=(a_1 \otimes I_{m_{a_2}}, I_{m_{a_1}} \otimes a_2)$ and  $H_Z=(I_{n_{a_1}} \otimes a_2^T, a_1^T \otimes I_{n_{a_2}})$. The resulting quantum code has a length of $n=n_{a_1}m_{a_2}+n_{a_2}m_{a_1}$ and a dimension of $k=2k_{a_1}k_{a_2}-k_{a_1}(n_{a_2}-m_{a_2})-k_{a_2}(n_{a_1}-m_{a_1})$. Here, we focus on GHP codes which contain both generalized bicycle codes~\cite{GBcodes} and the class of HP codes, where one of the two parity-check matrices used in the product is square~\cite{osd}. 

Our analysis is based on enumerating trapping sets that are children of the $g$ and ($g+2$)-cycles of the code. 
To search for the children trapping sets, we first check if the degree-$1$ check nodes of the parent trapping set share any variable nodes. 
If not, we proceed by checking if there are variable nodes shared between degree-$1$ and  degree-$2$ check nodes of the parent trapping set. The procedure continues until all degree-$1$ check nodes are checked against all check nodes of all possible degrees. 
We proceed similarly to the expansion procedure in~\cite{expansion}. The termination condition is when no degree-$1$ check nodes are left.

The $B1$ GHP code is used as a case study for the upcoming analysis. The $A$ and $B$ matrices that describe the $B1$ code are given below.
\[
A=
\begin{pmatrix}
x^{27} & 0 & 0 & 0 & 0 & 1 & x^{54}\\
x^{54} & x^{27} & 0 & 0 & 0 & 0 & 1\\
1&x^{54} & x^{27} & 0 & 0 & 0 & 0\\
0&1&x^{54} & x^{27} & 0 & 0 & 0\\
0&0&1&x^{54} & x^{27} & 0 & 0 \\
0&0&0&1&x^{54} & x^{27} & 0\\
0&0&0&0&1&x^{54} & x^{27}\\
\end{pmatrix},
\]
\[
B=(1+x+x^6)I_7.
\]
Regarding the $A$ circulant matrix, each entry corresponds to a $63 \times 63$ matrix ($l=63$), with entry $0$ representing the all-zero matrix, entry $1$ corresponding to the identity matrix and entries $x^{27}$ and $x^{54}$ being the shifted versions (by $27$ and $54$) of the identity matrix. Regarding the $B$ circulant matrix, each entry also corresponds to a $63 \times 63$ matrix which is repeated seven times across the diagonal of $B$ ($I_7$ corresponds to the $7\times 7$ identity matrix). 
For the $H_Z$ matrix, each entry of both $A$ and $B$ matrices is transposed~\cite{osd}. The code is regular, with $d_v=3$ and $d_c=6$ and its minimum distance is $18 \leq d \leq 24$.
Without loss of generality, since we only consider $X$ errors, our focus is on the $H_Z$ matrix of the code. For the upcoming analysis, we will refer to the two circulant matrices of $H_Z$ as $B^*$ and $A^*$. 

 \subsection{6-cycles and Classical Trapping Sets}     
\begin{figure}[h]
\centering
\begin{subfigure}[]
{   \centering 
\includegraphics[width=.065\textwidth]{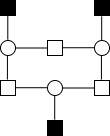}
\label{fig:33}
}
\end{subfigure}

\begin{subfigure}[]
{
\centering
\includegraphics[width=.12\textwidth]{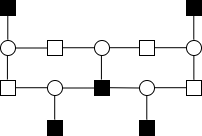}
\label{fig:55}
}
\end{subfigure}
\begin{subfigure}[]
{
\centering
\includegraphics[width=.12\textwidth]{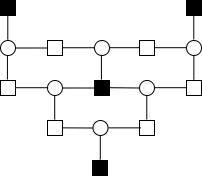}
\label{fig:64}
}
\end{subfigure}
\begin{subfigure}[]
{
\centering
\includegraphics[width=.17\textwidth]{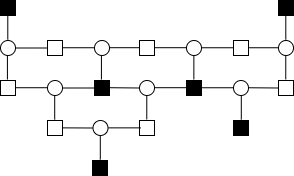}
\label{fig:86}
}
\end{subfigure}
\caption{Children trapping sets of the $(3,3)$ trapping set in~\subref{fig:33} are the $(5,5)$ trapping set in~\subref{fig:55}, the $(6,4)$ trapping set in~\subref{fig:64}, and the $(8,6)$ trapping set in~\subref{fig:86}. 
Each child of the $(3,3)$ trapping set is a juxtaposition of two, three, and four six-cycles, respectively. Odd-degree check nodes are represented by $\blacksquare$.}
\label{fig:childrenof3}
\end{figure}

Let us now consider the $6$-cycles of the code. We observed that all $6$-cycles are only formed between variable nodes that belong to the same circulant matrix. There are no $6$-cycles formed between variable nodes that belong to both circulant matrices. The parent trapping set which corresponds to a $6$-cycle is the $(3, 3)$ one, which is also a fixed set and can be seen in Fig.~\ref{fig:childrenof3}\subref{fig:33}.

\begin{figure}[htbp]
     \centering
     \begin{subfigure}[]
     {
        \centering
        \includegraphics[width=.13\textwidth]{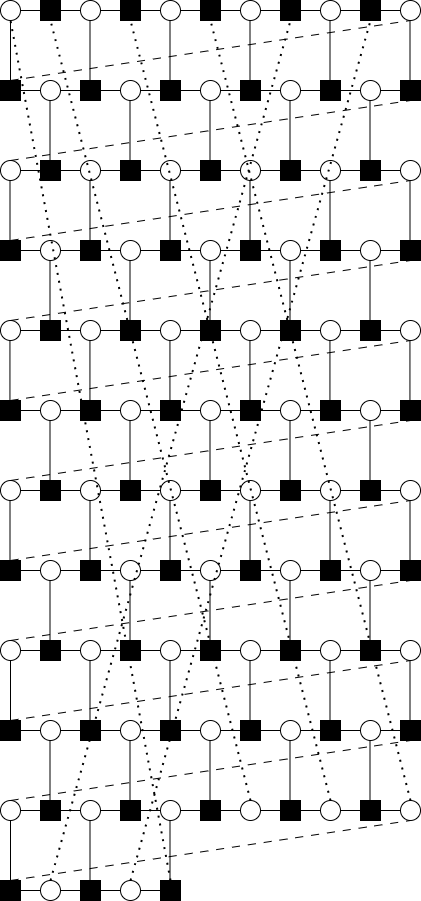}
         \label{fig:stopping0}
    }
    \end{subfigure}
    \begin{subfigure}[]
    {
        \centering
        \includegraphics[width=.12\textwidth]{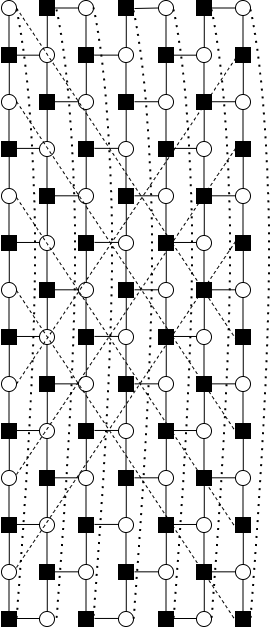}
        \label{fig:stopping1}
    }
    
    \end{subfigure}
    \caption{The $(63,63)$ and $(49,49)$ trapping sets obtained by the the $(3, 3)$ trapping set in the $B^*$  and $A^*$  circulant matrices respectively.  $B^*$ contains seven $(63,63)$ trapping sets and $A^*$ contains nine $(49,49)$ trapping sets. $(63,63)$ trapping sets are only interconnected with $(49,49)$ trapping sets and vice versa. Note that only degree-$3$ check nodes appear in the trapping sets. We use two styles of dotted lines to represent connections between remote nodes. The five $6$-cycles in a row, seen in~\subref{fig:stopping0}, can be inferred from the polynomial $1+x+x^6$.} 
    \label{fig:stopping}
\end{figure}

We start by considering the $(3,3)$ trapping set inside the $B^*$ matrix. 
By searching for children trapping sets, we obtain the structures shown in Fig.~\ref{fig:childrenof3}\subref{fig:55},\subref{fig:64},\subref{fig:86}, which correspond to interconnected $6$-cycles (note that \subref{fig:64} is also a fixed set). 
By enumerating the children of the $6$-cycle we eventually get the structure depicted in Fig.~\ref{fig:stopping}\subref{fig:stopping0}, which is a $(63, 63)$ trapping set. 
This structure corresponds to each $63\times 63$ entry of the diagonal of the $B^*$ matrix. It is clear that the $(63, 63)$ trapping set has a symmetric structure.
We note that the $B^*$ matrix consists of seven $(63, 63)$ trapping sets which are independent with respect to $B^*$, meaning that the checks of each $(63, 63)$ trapping set are not connected to variable nodes that belong to other $(63, 63)$ trapping sets. 

We now obtain a $6$-cycle of the $A^*$ matrix, which is also the parent to trapping sets formed by juxtaposing $6$-cycles. By searching for children trapping sets, we finally obtain the $(49, 49)$ trapping set which is shown in Fig.~\ref{fig:stopping}\subref{fig:stopping1}.
The $A^*$ circulant matrix is therefore comprised of nine $(49, 49)$ trapping sets which are not interconnected with each other and are only connected with the $(63, 63)$ trapping sets of the other circulant matrix. We stop our search here since the $(49,49)$ and $(63,63)$ trapping sets also constitute stopping sets~\cite{pseudo}, which are directly connected to pseudocodewords and are harmful when decoded over the BSC. Pseudocodewords are considered to be attractor points of iterative message passing decoders~\cite{ITW}, analogous to codewords, thereby determining the performance of iterative decoding~\cite{psDecoding}.

\subsection{8-cycles and Quantum Trapping Sets}
\begin{figure}
    \centering
    \begin{subfigure}[]
    {
        \centering
        \includegraphics[width=.15\textwidth]{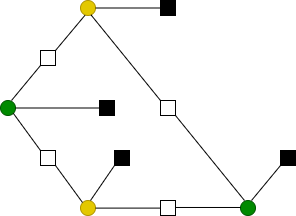}
        \label{fig:44}
        }
    \end{subfigure}
    \begin{subfigure}[]
    {
        \centering
        \includegraphics[width=.15\textwidth]{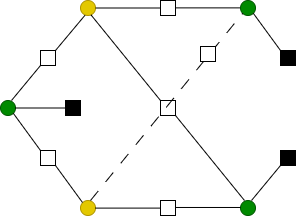}
        \label{fig:53}
        }
    \end{subfigure}
    \begin{subfigure}[]
    {
        \centering
        \includegraphics[width=.15\textwidth]{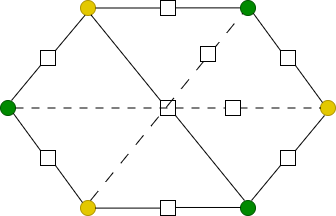}
        \label{fig:60TS}
        }
     \end{subfigure}
    \caption{The evolution of a $(4,4)$ trapping set~\subref{fig:44} leading into the formation of a symmetric stabilizer~\subref{fig:60TS}. The $(5,3)$ trapping set~\subref{fig:53} is obtained after adding a single variable node and the $(6,0)$ trapping set is obtained after adding two variable nodes. The three line types used for the $(6,0)$ trapping set indicate that the lines are not interconnected. Notice that only variable nodes belonging to different circulant matrices are directly connected.} 
    \label{fig:elemb}
\end{figure}

Let us now consider the $8$-cycles of the code. We observed that $8$-cycles are only formed between sets of variable nodes that belong to both circulant matrices and there is no $8$-cycle inside each circulant matrix independently. The parent trapping set corresponding to an $8$-cycle is the $(4,4)$ one which can be seen in Fig.~\ref{fig:elemb}\subref{fig:44}. Yellow-colored circles correspond to variable nodes of the $B^{*}$ circulant matrix, and green-colored variable nodes correspond to variable nodes of the $A^{*}$ circulant matrix. Now, we will try to connect the degree-$1$ check nodes by adding one variable node. Apparently, as illustrated in Fig.~\ref{fig:elemb}\subref{fig:44}, the degree-$1$ checks connected to yellow-colored variable nodes, share a neighboring variable node, thus forming the $(5,3)$ trapping set (Fig.~\ref{fig:elemb}\subref{fig:53}). Finally, all the dangling degree-$1$ checks are connected when we add one more node to the $(5,3)$ trapping set. This leads us to obtain the $(6,0)$ trapping set, which is a symmetric stabilizer and is illustrated in Fig.~\ref{fig:elemb}\subref{fig:60TS}. The $(6,0)$ trapping set consists of six interconnected $8$-cycles formed between the two circulant matrices.
Each variable node is contained in three $(6,0)$ trapping sets and $18$ $8$-cycles. Every $8$-cycle of the code leads to the formation of the quantum trapping set. The set of six variable nodes inside the $(6,0)$ trapping set corresponds to the support of a check of the $H_X$ matrix.

\begin{figure}[h]
    \centering
    \includegraphics[width=.4\textwidth]{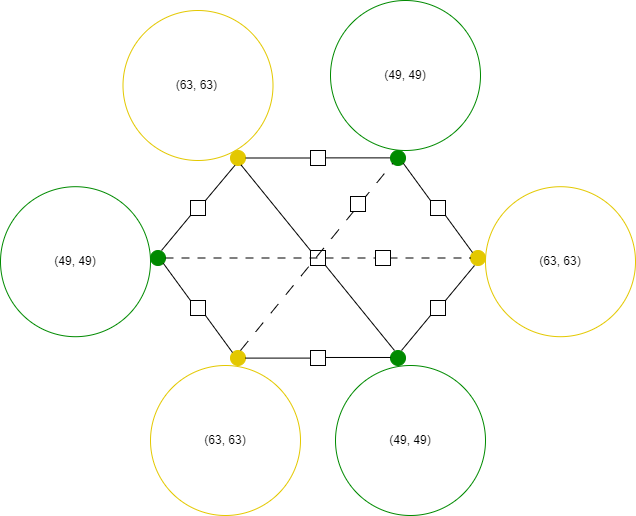}
    \caption{Overview of the structure of $B1$ code. The symmetric stabilizer interconnects the two types of trapping sets. The $(63,63)$ and $(49,49)$ trapping sets are represented as yellow and green circles respectively, which indicate their symmetric structure.}
    \label{fig:total}
\end{figure}

Also note that the $(6,0)$ trapping sets interconnect the $(63,63)$ and $(49,49)$ classical trapping sets as illustrated in Fig.~\ref{fig:total}. 

Finally, regarding every $d_v$-$3$, $d_c$-$6$ GHP code, we conjecture that since one of the circulant matrices will be constructed by using a weight-$3$ polynomial, then the trapping sets of that circulant matrix will be a juxtaposition of $6$-cycles. We also conjecture that the $(6,0)$ trapping set will be present in every $d_v$-$3$, $d_c$-$6$ GHP code. Hence, it is crucial to devise a decoding framework that is able to correct error patterns appearing in such structures.   

\section{Two-Bit Bit Flipping Decoding}
\label{sec:tbf}


The trapping set analysis of the $B1$ code indicates that BF decoder is expected to exhibit poor performance for $d_v$-$3$ GHP codes, despite being a good candidate for low-latency applications. In particular, it has been proven that for $d_v$-$3$ variable-regular codes, the guaranteed error correction capability of the BF algorithm is upper bounded by $\lceil \frac{g}{4} \rceil-1$~\cite{5437420}. 
Therefore, the fact that a $d_v$-$3$ code with $g = 6$ or $g = 8$ cannot correct certain weight two error patterns makes the algorithm impractical. 
Also, the girth of QLDPC codes is usually small, and in the case of HP and GHP codes, it cannot be higher than eight. 
Hence, we are motivated to utilize bit flipping-based decoders with better error correction capabilities, accompanied by higher granularity, and allowing collective decoding. 
\subsection{TBF Algorithm}
We now present the quantum equivalent of the classical two-bit bit flipping (TBF) algorithm~\cite{TBFA}. 
In contrast to the BF decoder, which assigns two possible states to each variable and check node (one-bit precision), TBF uses two-bit precision, allowing more states to describe both variable and check nodes. 

For the check nodes, an additional bit is used to describe the value of the check in the previous iteration. 
A satisfied (unsatisfied) check node $c$ is called \textit{previously satisfied} (\textit{previously unsatisfied}) if it was satisfied (unsatisfied) in the previous decoding iteration and it is still satisfied (unsatisfied) in the current iteration, which means that its residual syndrome $r_c=0$ (or $r_c=1$) for both iterations. 
In the other case, the check node is called \textit{newly satisfied} (\textit{newly unsatisfied}), which means that the residual syndrome $r_c$ flips in the current iteration. 
The states of a previously satisfied, a newly satisfied, a previously unsatisfied, and a newly unsatisfied check node are denoted as $0_{\text{old}}$, $0_{\text{new}}$, $1_{\text{old}}$, and $1_{\text{new}}$, respectively.

Let $\mathbf{z}^{\ell}=(z_1^\ell, z_2^\ell,\cdots,z_m^\ell)$ be a vector 
such that $z_c^\ell \in \{0_{\text{old}},0_{\text{new}},1_{\text{old}},1_{\text{new}}\}$ gives the state of check node $c$ at the end of the $\ell^{\text{th}}$ iteration. 
The check node update function $\Phi$ is defined as follows: 
\begin{equation}
z_c^{\ell} \gets \Phi(r_c^{\ell-1},r_c^{\ell})= \begin{cases}
			 0_{\text{old}}, & \text{if $ (r_c^{\ell-1},r_c^{\ell})=(0,0)$}\\
            1_{\text{new}}, & \text{if $(r_c^{\ell-1},r_c^{\ell})=(0,1)$}\\
             0_{\text{new}}, & \text{if $(r_c^{\ell-1},r_c^{\ell})=(1,0)$}\\
            1_{\text{old}}, & \text{if $(r_c^{\ell-1}, r_c^{\ell})=(1,1)$}.
		 \end{cases}
    \label{eq:phi}
\end{equation}
The initial states of check nodes are computed at iteration $\ell = 0$ based on the received syndrome as follows:
\begin{equation}
z_c^{0} \gets \Delta_c(s_c)= 
\begin{cases}
        \{0_{\text{old}}, 0_{\text{new}}\}, & \text{if $s_c=0$}\\
   \{1_{\text{old}}, 1_{\text{new}}\}, & \text{if $s_c=1$}.
\end{cases} 
\label{eq:initChecks}
\end{equation}
Hence, if the received syndrome of a check equals zero, the check can be initialized either as previously satisfied or as newly satisfied. On the other hand, if the received syndrome of a check equals one, the check can be initialized either as previously unsatisfied or as newly unsatisfied. This is an indication of the granularity offered by TBF decoding, which can be exploited via collective decoding.

In the case of variable nodes, the additional bit is used to represent their ``strength". A variable node $v$ is classified as either a ``strong" variable node or a ``weak" variable node. The intuition behind this classification is that variable nodes are treated differently with respect to their strength. For example, a variable node that is marked as weak can be flipped more easily than a variable node marked as strong in the subsequent iteration. 
Therefore, $v$ can be described by four states instead of two, which are the following: $00, 01, 10, 11$, where the most significant bit (MSB) denotes the value of $v$ and the least significant bit (LSB) denotes the strength of $v$. 

Let $\mathbf{w}^{\ell}=(w_1^\ell, w_2^\ell,\cdots,w_n^\ell)$, such that $w_v^\ell\in \{00,01,10,11\}$ gives the state of variable node $v$ at the end of the $\ell^{\text{th}}$ iteration.
The hard decision of $v$  at the $\ell^{\text{th}}$ iteration is extracted using the MSB as: 
$\text{MSB}(w_v^{\ell})$ and the error estimate vector $\mathbf{\hat{e}}^{\ell}$ corresponds to the hard decision part of $\mathbf{w}^{\ell}$.

Let us spend some time understanding how the variable nodes are updated in the TBF algorithm. States of variable nodes are initialized to $\Delta_v \in \{00, 01\}$$ $ since we do not receive any channel values. 
The variable node update function used to compute the updated variable node state is defined by \eqref{eq:f2}: 
\begin{equation}
w_v^{\ell+1}=\begin{cases}
			w_v^{\ell}, & \text{if $\mathcal{X}^{\ell}(v) = (2,0,0)$}\\
            00, & \text{if $\mathcal{X}^{\ell}(v) = (1,0,1)~\&~ w_v^{\ell} \in (01,00)$}\\
            10, & \text{if $\mathcal{X}^{\ell}(v) = (1,0,1)~\&~ w_v^{\ell} \in (10,11)$}\\
            \Psi(w_v^{\ell},\chi_1^{\ell}(v)), & \text{otherwise}.
		 \end{cases}
\label{eq:f2}
\end{equation}
Here, $\mathcal{X}^{\ell}(v)=(\chi_{0_\text{old}}^\ell(v)$, $\chi_{0_\text{new}}^\ell(v)$, $\chi_{1_\text{old}}^\ell(v))$ denotes the 3-tuple containing the \emph{number} of previously satisfied, newly satisfied, and previously unsatisfied neighboring check nodes of $v$ respectively, and the function $\Psi$ is described in Table~\ref{tab:f1}. 
The function $\Psi$ computes the next state of each variable node by using the number of its adjacent unsatisfied checks $\chi_1^{\ell}(v)$ and its current state $w_v^{\ell}$. In Table~\ref{tab:f1}, for the case of three unsatisfied checks, both weak and strong bits are flipped to strong bits. In the case of two unsatisfied checks, strong bits become weak and, in turn, weak bits are flipped to strong bits. Finally, for one unsatisfied check, weak bits are flipped to weak bits, whereas for no unsatisfied checks, bits become strong while retaining their hard decision value. Note that $\Psi$ defines~(\ref{eq:f2}) for any input configuration except the following two cases. 
Firstly, when a variable node $v$ is connected to two previously satisfied checks, zero newly satisfied checks, and zero previously unsatisfied check nodes - i.e., $\mathcal{X}^{\ell}(v) =(2,0,0)$, then, we do not modify the state of $v$. Secondly, when
$\mathcal{X}^{\ell}(v) =(1,0,1)$, the decoding rule forces $v$ to become weak, i.e., LSB is set to zero. All such subtleties can be captured in the variable node update function in a TBF algorithm.

\begin{table}[h]
\caption{Description of $\Psi$ which assigns the state of a variable node $v$ at the $(\ell+1)^{\text{th}}$ iteration based on its current state $w_v^\ell$ and the number of its current adjacent unsatisfied checks $\chi_1^{\ell}(v)$.}
\label{tab:f1}
\centering
\begin{tabular}{cc|c|c|c|}
\hline
\multicolumn{1}{|c|}{\multirow{2}{*}{$w_v^{\ell}$}}       
& \multicolumn{4}{c|}{$\chi_1^{\ell}(v)$} \\ \cline{2-5}
\multicolumn{1}{|c|}{}                                 
        & $0$ & $1$ &   $2$ & $3$\\ \hline
        \multicolumn{1}{|c|}{\begin{tabular}[c]{@{}c@{}}$01$\end{tabular}}   
         & $01$  & $01$ & $00$ & $11$ \\ \hline
        \multicolumn{1}{|c|}{\begin{tabular}[c]{@{}c@{}}$00$\end{tabular}} 
         & $01$  & $10$ & $11$ & $11$ \\ \hline
        \multicolumn{1}{|c|}{\begin{tabular}[c]{@{}c@{}}$11$\end{tabular}} 
        & $11$ & $11$ & $10$ & $01$   \\ \hline
        \multicolumn{1}{|c|}{\begin{tabular}[c]{@{}c@{}}$10$\end{tabular}} 
        & $11$ & $00$ & $01$ & $01$    \\ \hline
    \end{tabular}
\end{table}

We note that~\eqref{eq:f2} and $\Psi$ function of Table~\ref{tab:f1} correspond to a single example of a realization of a TBF decoder. More generally, the variable node update can have many different realizations giving more flexibility in the decoding rule. We will use multiple realizations of~(\ref{eq:f2}) in order to develop our collective decoding framework. Considering the definitions given above, the syndrome-based TBF decoder is described in Algorithm~\ref{alg:tbfa}. 
\begin{algorithm}
    \caption{Syndrome-Based TBF Algorithm}
     \label{alg:tbfa}
  \textbf{Input}: $\mathbf{s}, H$, $L$ 

  \textbf{Output}: $\hat{\mathbf{e}}$

  \textbf{Parameters}: $\Phi, \Psi$
  \begin{algorithmic}[1]
      \State $\ell\gets 0$ \Comment{Initializations}
      \State $\hat{e}_v^{\ell} \gets 0$, $\forall v \in \mathcal{V}$ 
      \State $w_v^{\ell}\gets \Delta_v$, $\forall v \in\mathcal{V}$
      \State $z_c^{\ell}\gets \Delta_c({s_c})$, $\forall c \in \mathcal{C}$
      \State $\mathbf{r}^{\ell} \gets \mathbf{s}$
      \While{$\mathbf{r}^{\ell} \neq 0$ and $\ell \leq L$}
      \State $\forall v:$ $w_v^{\ell+1} \gets 
      (\ref{eq:f2})$
       \State  $\hat{e}_v^{\ell+1} \gets \text{MSB}(w_v^{\ell+1})$, $\forall v \in \mathcal{V}$ \Comment{Hard decision of $w_v^{\ell}$}
         \State $\hat{s}^{\ell+1}=\mathbf{\hat{e}}^{\ell+1}H^T(\text{mod}2)$ \Comment{Estimated syndrome}
      \State  $r_c^{\ell+1}=s_c\xor\hat{s}_c^{\ell+1}$, $\forall c \in \mathcal{C}$ \Comment{Residual syndrome}
       \State $z_c^{\ell+1} \gets \Phi(r_c^{\ell},r_c^{\ell+1})$, $\forall c \in \mathcal{C}$
        \State $\ell\gets\ell+1$
      \EndWhile
  \end{algorithmic}
\end{algorithm}

\subsection{Designing TBF Decoders}
The granularity offered by the variable node update function~(\ref{eq:f2}) can help us develop TBF decoders with diverse error-correction capabilities.
Each TBF decoder can be uniquely determined as:
$$D =(\Psi,f),$$ where $f$ is a function that determines how the variable nodes are updated as described next. 
Variable node function $\Psi$ can be described as a look-up-table like the one shown in Table~\ref{tab:f1}, and in general, it can be applied to any subset of variable nodes. Multiple $\Psi$ functions can be deployed to cover the whole set of variable nodes $\mathcal{V}$. In such case, a $\Psi$ function corresponding to a subset $\mathcal{V}_i \subseteq \mathcal{V}$ will be denoted as $\Psi^{\mathcal{V}_i}$. For the majority of decoders we used in our simulations, $\Psi$ is as described in Table~\ref{tab:f1}. This is based on the observation that one should be cautious while relaxing the bit flipping criteria of the $\Psi$ function, as logical errors can be introduced in cases where bits with small number of unsatisfied checks are simultaneously flipped. Consequently, most of our decoders are defined by modifying the $f$ function, which makes the bit flipping procedure more conservative.

We define $f$ to be a sequence of $10$ binary values which determines how~(\ref{eq:f2}) is applied to any subset $\mathcal{V}_i \subseteq \mathcal{V}$ of variable nodes, as: \begin{multline}f^{\mathcal{V}_i}=(\mathcal{I}_{\Delta_v},\mathcal{I}_{\Delta_c},W_{012},W_{120},W_{200},W_{201},W_{101},W_{021}\\ W_{011},W_{020}).\end{multline}
The first two parameters act on the initialization of variable and check nodes, i.e., $\Delta_v$ and $\Delta_c$, respectively.
If $\mathcal{I}_{\Delta_v}=1$, then all variable nodes are initialized as $\Delta_v=00$ (weak zero), otherwise, they are initialized as $\Delta_v=01$ (strong zero). 
If $\mathcal{I}_{\Delta_c}=1$, then check nodes are initialized as $\Delta_c(s_c)=\{0_\text{new},1_\text{new}\}$ (newly satisfied/unsatisfied), depending on their value, $s_c$. If $\mathcal{I}_{\Delta_c}=0$, check nodes are initialized as $\Delta_c(s_c)=\{0_\text{old},1_\text{old}\}$ (previously satisfied/unsatisfied) depending on their value, $s_c$. For the rest of the parameters, the subscript $3$-tuple corresponds to $\mathcal{X}^{\ell}(v)$, where $v \in \mathcal{V}_i$. 
Now we will describe the effect of each parameter in the variable-to-check update rule.

If $W_{012}=1$, then the state of a variable node $v$ connected to zero previously satisfied, one newly satisfied, and two previously unsatisfied check remains the same for that iteration, thus: \begin{equation}
w_v^{\ell+1}=\begin{cases}
w_v^{\ell}, & \text{if ($W_{012}=1$~\&~$\mathbbm{1}_{012}(\mathcal{X}^{\ell}(v))=1$)}\\
\Psi(w_v^{\ell},\chi_1^{\ell}(v)), & \text{if ($W_{012}=0$~\&~$\mathbbm{1}_{012}(\mathcal{X}^{\ell}(v))=1$)}.
		 \end{cases}
\label{eq:f012}
\end{equation}
Where, \begin{equation}\mathbbm{1}_{012}(\mathcal{X}^{\ell}(v))= 
\begin{cases}
        1, & \text{if $\mathcal{X}^{\ell}(v)=(0,1,2)$}\\
   0, & \text{otherwise,}
\end{cases}\end{equation} 
is the indicator function. The remaining parameters are also described using equations similar to~(\ref{eq:f012}).
If $W_{120}=1$, then the state of a variable node connected to one previously satisfied, two newly satisfied, and zero previously unsatisfied check becomes weak, otherwise, if $W_{120}=0$, the state of the particular variable node remains the same for that iteration. This is described by the following equation:
\begin{equation}
w_v^{\ell+1}=\begin{cases}
00, & \text{if ($W_{120}=1$~\&~$\mathbbm{1}_{120}  (\mathcal{X}^{\ell}(v))=1$)} \\ & \&~w_v^{\ell} \in (01,00)\\
10, & \text{if ($W_{120}=1$~\&~$\mathbbm{1}_{120}  (\mathcal{X}^{\ell}(v))=1$)} \\ & \&~w_v^{\ell} \in (10,11)\\
w_v^{\ell}, & \text{if ($W_{120}=0$~\&~$\mathbbm{1}_{120}  (\mathcal{X}^{\ell}(v))=1$)}.
		 \end{cases}
\end{equation}
The same rule applies for $W_{200}$. Finally, if $W_{201}=1$, then the state of a variable node connected to two previously satisfied, zero newly satisfied, and one previously unsatisfied checks becomes weak. This is described by the following equation:
\begin{equation}
w_v^{\ell+1}=\begin{cases}
00, & \text{if ($W_{201}=1$~\&~$\mathbbm{1}_{201}  (\mathcal{X}^{\ell}(v))=1$)} \\ & \&~w_v^{\ell} \in (01,00)\\
10, & \text{if ($W_{201}=1$~\&~$\mathbbm{1}_{201}  (\mathcal{X}^{\ell}(v))=1$)} \\ & \&~w_v^{\ell} \in (10,11)\\
\Psi(w_v^{\ell},\chi_1^{\ell}(v)), & \text{if ($W_{201}=0$~\&~$\mathbbm{1}_{201}  (\mathcal{X}^{\ell}(v))=1$)}.
		 \end{cases}
\end{equation}
The rest of the parameters have the same behavior as $W_{201}$. 

In the following section, we will exploit the structure of GHP codes, analyzed in Section~\ref{sec:structure}, by using multiple instances of TBF decoders in parallel (collective decoding).

\section{Collective decoding}
\label{sec:dec}
Our collective decoding framework is denoted by $$\mathcal{D}_{N_{\mathcal{D}}}=\{D_i| i=1,\cdots, N_{\mathcal{D}}\},$$ which is the set of TBF decoders we have at our disposal, where $N_{\mathcal{D}}$ is the cardinality of the set and $D_i=(\Psi_i, f_i)$ describes each individual decoder of the set, which is uniquely defined.  
The set of $N_{\mathcal{D}}$ decoders is used in parallel so that it can provide guaranteed correction within the allowed decoding time budget. 
For the rest of the paper, if no superscript is defined for $\Psi$ and $f$ functions then we assume that they are applied to the whole set of variable nodes, $\mathcal{V}$. Regarding the $f$ function, although the number of all possible $3$-tuples is more than $10$, for the purposes of our decoding analysis and in order to limit the complexity, we choose to work with the aforementioned $10$ bit-values. We denote as $\mathcal{D}_{tot}$, the base of $2^{10}=1024$ decoders that we will have at our disposal (maximum value of $N_{\mathcal{D}}$ equals to $1024$). Most of our proposed decoders will use the $\Psi$ function defined in Table~\ref{tab:f1}. 
      
\subsection{Targeting Error-Floor}
Based on the trapping set analysis of the code in Section~\ref{sec:structure}, 
our goal is to correct all error patterns that appear inside the $(6,0)$ trapping set and all error patterns up to weight-$t$ inside the $(49, 49)$ and $(63, 63)$ trapping sets. Since those structures repeat themselves, the error-floor performance will be improved. We note here that we use the BF algorithm as a starting point of our analysis given below, because BF is a special case of a TBF decoder.

\subsubsection{Dealing with classical trapping sets}
\begin{figure}[htbp]
\centering
\begin{subfigure}[]
{
\centering
\includegraphics[width=.055\textwidth]{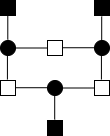}
\label{fig:threeA}
}
\end{subfigure}
\begin{subfigure}[]
{
\centering
\includegraphics[width=.1\textwidth]{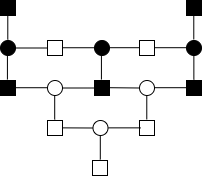}
\label{fig:threeB}
}
\end{subfigure}
\begin{subfigure}[]
{
\centering
\includegraphics[width=.1\textwidth]{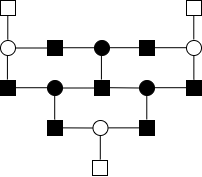}
\label{fig:threeC}
}
\end{subfigure}
\begin{subfigure}[]
{
\centering
\includegraphics[width=.14\textwidth]{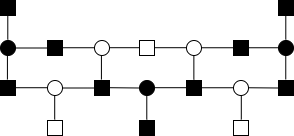}
 \label{fig:threeD}
         }
     \end{subfigure}
     \begin{subfigure}[]
     {
               \centering
    \includegraphics[width=.1\textwidth]{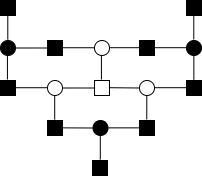}
\label{fig:threeE}
         }
     \end{subfigure}
      \caption{Uncorrectable weight-$3$ error configurations by BF decoder. These errors appear in smaller trapping sets within the $(63,63)$ and $(49,49)$ trapping sets. Variable nodes in error are marked with {\LARGE$\bullet$}. Note that in this case, $\blacksquare$ corresponds to an unsatisfied check.} 
     \label{fig:fail3}
\end{figure}

Since both $(49,49)$ and $(63,63)$ trapping sets are formed by juxtaposing $6$-cycles, the following analysis applies to both of them. We begin our analysis by considering BF, which flips every bit that has two or more unsatisfied checks in their neighborhood. BF can correct every weight-$1$ and weight-$2$ error pattern that appears in both $(49,49)$ and $(63,63)$ trapping sets. 
We now consider weight-$3$ error patterns. BF cannot correct the types of weight-$3$ error patterns that are depicted in Fig.~\ref{fig:fail3}. The weight-$3$ error patterns appearing inside the $(3, 3)$ trapping set (Fig.~\ref{fig:fail3}\subref{fig:threeA}) constitute fixed sets for BF decoder (see Section~\ref{sec:tbf}), so the decoder fails to correct them because the number of unsatisfied checks of all bits equals to one, therefore no variable node is flipped. Regarding the error patterns shown in Fig.~\ref{fig:fail3}\subref{fig:threeB},\subref{fig:threeC},\subref{fig:threeD},\subref{fig:threeE}, BF exhibits oscillating behavior as it does not provide different treatment for variable nodes with two unsatisfied checks and variable nodes with three unsatisfied checks. All these error patterns can be corrected by using the following TBF-based decoder which is more cautious while flipping bits: $D_1=(\Psi_1,(0,1,0,0,0,1,1,0,1,0))$, where $\Psi_1$ is given by Table~\ref{tab:f1}. $D_1$ can correct the error patterns of Fig.~\ref{fig:fail3} in five, four, one, one, and one iterations, respectively. 

Regarding weight-$4$ error patterns, we continue our analysis using the $D_1$ decoder as our basis. Our goal now is to first generate all possible weight-$4$ error patterns that can appear in the $(49,49)$ trapping set. For every weight-$4$ error pattern for which $D_1$ fails, we run the rest  $1023$ decoders, $\mathcal{D}_{tot}\backslash D_1$. Eventually, we pick a decoder of that set which is able to correct the largest number of weight-$4$ error patterns for which $D_1$ fails. It turns out that $D_2=(\Psi_2,(0,0,0,0,0,0,0,0,0,0))$, where $\Psi_2$ is given by Table~\ref{tab:f1}, is able to correct every single one of those error patterns. Now that we can correct all weight-$4$ error patterns inside the $(49,49)$ trapping set, we use both $D_1$ and $D_2$ decoders as our basis for testing the correction of all weight-$4$ error patterns appearing in the $(69,69)$ trapping set. Following the same approach as before, we found that decoder $D_3=(\Psi_3,(0,0,0,0,1,0,0,0,0,0))$, where $\Psi_3$ is given by Table~\ref{tab:f1}, can complement $D_1$ and $D_2$ decoders in correcting all possible weight-$4$ error patterns appearing in both $(49,49)$ and $(63,63)$ trapping sets.

Finally, Algorithm~\ref{alg:div} outlines a recursive approach that generates a set of TBF-based decoders that can collectively correct all error patterns up to a certain weight, which appear inside classical trapping sets. 
$\mathcal{E}^{j}$ denotes the set of all error patterns of weight-$j$ that are found inside the trapping set, while $t$ denotes the desired error correction capability inside the trapping set (we aim to correct all error patterns up to weight-$t$ in the trapping set). Also, $\mathcal{E}_D^{j}$, denotes the set of error patterns of weight-$j$ that are correctable by the decoder $D$. All decoders of the set $\mathcal{D}_{tot}$ are able to correct all weight-$1$ error patterns, so the algorithm begins by initializing the set of available decoders as the $D_1$ decoder. Then, for all error patterns of weight-$2$, the rest of all possible $1023$ decoders of the set, $\mathcal{D}_{tot}\backslash D_1$, are simulated and the one with the lowest frame error rate (FER), $D$, is added to the set $\mathcal{D}$. The error patterns that are correctable by $D$ are removed from the set $\mathcal{E}^{j}$. The process continues until either all error patterns of weight $j$ are corrected or all the decoders of the set $\mathcal{D}_{tot}\backslash D_1$ have been used. After all error patterns of a certain weight are corrected, the weight is incremented and the process continues. Eventually, Algorithm~\ref{alg:div} outputs the set of decoders $\mathcal{D}$ that can correct up to weight-($j-1$) error patterns appearing inside a trapping set. 

In the case of the $B1$ code, since both the $(49,49)$ and $(63,63)$ trapping sets are symmetric, we can reduce the cardinality of the error patterns of a certain weight. For example $|\mathcal{E}^{j}|=\binom{62}{j-1} + \binom{48}{j-1}$, which for $j=5$ equals $752425$. In general, the complexity of Algorithm~\ref{alg:div} depends on the number of variable nodes contained in the trapping sets under test and the parameter $t$.

\begin{algorithm}
    \caption{Generation of  Set $\mathcal{D}$} 
     \label{alg:div}
  \textbf{Input}:$t$, $\mathcal{D}_{tot}$, $\mathcal{E}^{2},\cdots, \mathcal{E}^{t}$ 

  \textbf{Output}: $\mathcal{D}, j-1$

  \begin{algorithmic}[1]
      \State $j\gets 2$\Comment{Initialization}
      \State $\mathcal{D} \gets D_1$ 
      \State $\mathcal{D}_{tot} \gets \mathcal{D}_{tot} \backslash D_1$ \Comment{Remove selected decoder}
      \State $FER(i) \gets 0$,  $\forall i \in [1,1024]$
      \While{$j \leq t$}
        \While{($\mathcal{E}^{j} \neq \emptyset$ \& $\mathcal{D}_{tot} \neq \emptyset$) }
        \State $\mathcal{E} \gets \emptyset$ 
      \For {$\forall \mathbf{e} \in \mathcal{E}^{j}$} 
      \State Run $\mathcal{D}$
      \If{$\mathcal{D}$ \text{fails}}
      \State $\mathcal{E} \gets \mathcal{E} \cup \mathbf{e}$ \Comment{Uncorrectable error patterns.}
      
      \EndIf
      \EndFor
      \If{$\mathcal{E} \neq \emptyset$}
      \For {$\forall \mathbf{e} \in \mathcal{E}$} 
       \State Run $\forall D_i \in \mathcal{D}_{tot}$
      \State store $FER(i)$ 
      \EndFor
      \State $D=\argmin_{D_i \in \mathcal{D}_{tot}}FER$
      \State $\mathcal{D} \gets \mathcal{D} \bigcup D$ \Comment{Include to the set of decoders}
      \State $\mathcal{E}^{j} \gets \mathcal{E}^{j} \backslash \mathcal{E}^{j}_D$ \Comment{ Correctable error patterns}
      \State $\mathcal{D}_{tot} \gets \mathcal{D}_{tot} \backslash D$ \Comment{Remove selected decoder}
      \Else
      
      ~~~~~~~$\mathcal{E}^{j} \gets \emptyset$
      \EndIf
      \EndWhile
      \State $j \gets j+1$
      \EndWhile
  \end{algorithmic}
\end{algorithm}

In Table~\ref{tab:div}, we list a set of eight TBF decoders that guarantee the correction of all error patterns up to weight-$5$ that appear inside the $(49,49)$ and $(63,63)$ trapping sets. Simulation results obtained by using this set of decoders are shown in Section~\ref{sec:sim}.

 \begin{table}[]
    \caption{Set of eight TBF decoders which guarantee the correction of all error patterns up to weight-$5$ that appear inside the $(49, 49)$ and $(63, 63)$ trapping sets of the $B1$ code (children of the $(3, 3)$ trapping set). The set of decoders is generated using Algorithm~\ref{alg:div}.}
    \label{tab:div}
    \centering
    \begin{tabular}{|c|c|c|}
           \hline
           Decoder  & $ (\scriptstyle \mathcal{I}_{\Delta_v},\mathcal{I}_{\Delta_c},W_{012},W_{120},W_{200},W_{201},W_{101},W_{021},W_{011},W_{020})$\\
           \hline
           $D_1$  & $(0,1,0,0,0,1,1,0,1,0)$ \\
            \hline
            $D_2$ & $(0,0,0,0,0,0,0,0,0,0)$\\ 
            \hline
            $D_3$  & $(0,0,0,0,1,0,0,0,0,0)$\\
            \hline
             $D_4$  & $(0,0,0,0,0,1,0,0,0,0)$\\
            \hline
             $D_5$  & $(1,1,0,0,0,0,0,0,1,1)$\\
            \hline
             $D_6$  & $(0,0,0,1,0,0,0,0,0,1)$\\
            \hline
             $D_7$  & $(1,1,0,0,0,0,1,1,0,0)$\\
            \hline
             $D_8$  & $(0,1,0,0,0,1,0,1,1,1)$\\
            \hline
    \end{tabular}
\end{table}

\subsubsection{Dealing with quantum trapping sets} 

\begin{figure}[tbh]
    \centering
    \includegraphics[width=.25\textwidth]{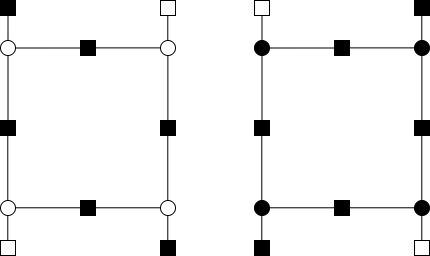}
    \caption{A weight-$2$ error pattern corresponding to variable nodes of the diagonal of the $(4,4)$ trapping set. BF oscillates between the two configurations, by flipping all four variable nodes at every iteration.}
    \label{fig:diagonal}
\end{figure}

We consider again the plain BF algorithm. It is easy to see that all weight-$1$ error patterns appearing inside the $(6,0)$ trapping set can be corrected by BF. However, when it comes to the weight-$2$ error patterns, BF can correct all of them except the ones that appear in diagonal positions of each $8$-cycle, or $(4,4)$ trapping set. As we observe in Fig.~\ref{fig:diagonal}, based on the decoding rule, BF keeps flipping all four nodes back and forth without treating differently the variable nodes with two or three unsatisfied checks. Hence, there are $\binom{4}{2}=6$ failing weight-$2$ error patterns per $(4, 4)$ trapping set. This problem can be solved by using a less ``aggressive" BF algorithm that introduces more granularity, such as $D_1$. By observing the parameters of $D_1$, it is clear that it will immediately flip all variable nodes with three unsatisfied checks while it will flag the variable nodes with two unsatisfied checks as ``weak", so it will be able to correct such error configurations in one iteration. 
      
\begin{figure}
    \centering
    \begin{subfigure}[]
    {
        \centering
        \includegraphics[width=.15\textwidth]{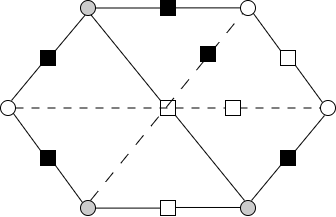}
        \label{fig:60A}
    }
    \end{subfigure}
    \begin{subfigure}[]
    {
        \centering
        \includegraphics[width=.15\textwidth]{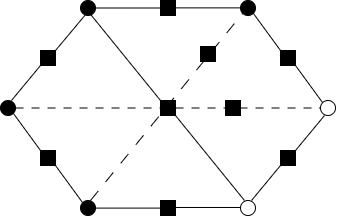}
        \label{fig:60B}
    }
    \end{subfigure}
    \begin{subfigure}[]
    {
        \centering
        \includegraphics[width=.15\textwidth]{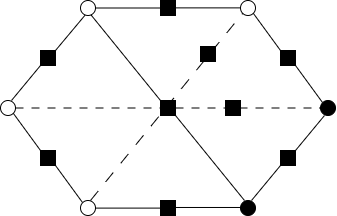}
        \label{fig:60C}
    }
    \end{subfigure}
    \caption{Progression of BF algorithm for a weight-$3$ error pattern appearing inside the $(6, 0)$ trapping set.  \subref{fig:60A} represents the initial state of the decoder, and the error pattern is marked with gray. After every node with two unsatisfied checks has been flipped, state~\subref{fig:60B} is obtained, and then BF oscillates between states~\subref{fig:60B} and \subref{fig:60C} as it always flips the same number of nodes belonging to the two different circulants. Such an error configuration can be resolved by treating nodes of different circulant matrices differently, thus inducing asymmetry.}
    \label{fig:symmetric}
\end{figure}

\begin{figure}
    \centering
    \begin{subfigure}[]
    {
        \centering
        \includegraphics[width=.15\textwidth]{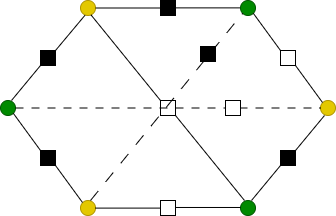}
        \label{fig:60A1}
    }
    \end{subfigure}
    \begin{subfigure}[]
    {
        \centering
        \includegraphics[width=.15\textwidth]{Figures/error60b.png}
        \label{fig:60B1}
    }
    \end{subfigure}
    \begin{subfigure}[]
    {
        \centering
        \includegraphics[width=.15\textwidth]{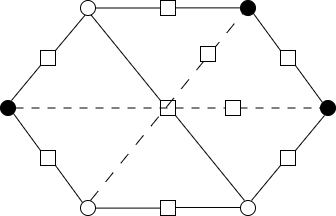}
        \label{fig:60C1}
    }
    \end{subfigure}
    \caption{Same error configuration depicted in Fig.~\ref{fig:symmetric} is now handled using a single TBF-based decoder ($D_9$) which employs different rules for variable nodes of each circulant matrix. Yellow and green colors differentiate the variable nodes belonging to different circulant matrices. As we observe nodes in green, despite having three unsatisfied checks, are not immediately flipped by $D_9$, hence the decoder is able to converge to a degenerate error pattern (which is equivalent to the gray one in Fig.~\ref{fig:symmetric}).}
    \label{fig:symmetricTBFA}
\end{figure}

We continue our analysis by considering weight-$3$ error patterns, all of which are uncorrectable by $D_1$ due to the symmetry of the $(6,0)$ trapping set. Since $(6, 0)$ trapping sets are only formed between the two circulant matrices, we can deploy a TBF decoder that assigns different rules to variable nodes regarding which circulant matrix they belong to. For example, variable nodes with three unsatisfied checks in one circulant are immediately flipped, while variable nodes with three unsatisfied checks in the other circulant only get their energy reduced (become ``weak") so they can potentially be flipped in subsequent iterations. This action introduces asymmetry and the overall decoder introduces a ``timing offset" to the variable nodes of one circulant matrix. An example of a weight-$3$ error configuration that cannot be corrected by either BF or $D_1$ is shown in Fig.~\ref{fig:symmetric}. Fig.~\ref{fig:symmetricTBFA} shows how such an error configuration can be corrected by employing different rules in each circulant matrix.

In conclusion, regarding the case of quantum trapping sets, we propose a single TBF algorithm that is more conservative in flipping the variable nodes of one circulant matrix and more aggressive with the variable nodes of the other circulant matrix. This approach enables the algorithm to handle all possible error patterns in the $(6,0)$ trapping set. We denote that algorithm as 
$$D_9=(\{\Psi_9^{\mathcal{V}_1}, \Psi_9^{\mathcal{V}_2}\}, (0,1,0,0,0,1,1,0,1,0)),$$
where $\Psi_9^{\mathcal{V}_1}$ is described by Table~\ref{tab:f1}, $\Psi_9^{\mathcal{V}_2}$ is described by Table~\ref{tab:f11}, $\mathcal{V}_1=\{v_1, v_2,..., v_{n/2}\}$ and $\mathcal{V}_2=\{v_{n/2+1}, v_{n/2+2}, ..., v_n\}$. Due to its inherent asymmetry, $D_9$ can correct every error pattern inside the $(6, 0)$ trapping set.

Finally, from this analysis it is clear that the set $\mathcal{D}_9=\{D_1,D_2,D_3,D_4,D_5,D_6,D_7,D_8,D_9\}$ is able to correct any error pattern up to weight-$5$ appearing inside the $(49,49)$ and $(63,63)$ classical trapping sets and any error pattern appearing inside a $(6,0)$ quantum trapping set.

\begin{table}[h]
    \caption{A realization of the $\Psi$ function, which is only applied on variable nodes of one circulant matrix and in conjunction with the update rule described by Table~\ref{tab:f1} corrects error patterns inside quantum trapping sets.\label{tab:f11}}
    \centering
    \begin{tabular}{cc|c|c|c|}
        \hline
        \multicolumn{1}{|c|}{\multirow{2}{*}{$w_v^{\ell}$}} & \multicolumn{4}{c|}{$\chi_1^{\ell}(v)$} \\ \cline{2-5}
        \multicolumn{1}{|c|}{} & $0$ & $1$ & $2$ & $3$ \\ \hline
        \multicolumn{1}{|c|}{\begin{tabular}[c]{@{}c@{}}$01$\end{tabular}} & $01$ & $01$ & $00$ & $00$ \\ \hline
        \multicolumn{1}{|c|}{\begin{tabular}[c]{@{}c@{}}$00$\end{tabular}} & $01$ & $10$ & $11$ & $11$ \\ \hline
        \multicolumn{1}{|c|}{\begin{tabular}[c]{@{}c@{}}$11$\end{tabular}} & $11$ & $11$ & $10$ & $10$ \\ \hline
        \multicolumn{1}{|c|}{\begin{tabular}[c]{@{}c@{}}$10$\end{tabular}} & $11$ & $00$ & $01$ & $01$ \\ \hline
    \end{tabular}
\end{table}

\subsection{Further Optimizations}
      \label{sec:decb}
The analysis of the $B1$ code in Section~\ref{sec:structure} revealed that the code is comprised of nine $(49, 49)$ trapping sets and seven $(63, 63)$ trapping sets which are connected via symmetric stabilizers. We can correct higher-weight error patterns by assigning different decoding rules to variable nodes depending on which circulant matrix they belong to. The type of error patterns we aim to correct are comprised of up to weight-$5$ error-patterns appearing inside classical trapping sets of one circulant matrix and error patterns appearing inside a symmetric stabilizer (quantum trapping set). For example, we observe that $D_1$ and $D_9$ are comprised of exactly the same function $f$ with the difference that $D_9$ assigns a different $\Psi$ function to the variable nodes of the second circulant matrix. This observation means that the set $\{D_1,D_9\}$ can guarantee the correction of all error patterns up to weight-$3$, which appear inside any of the classical trapping sets of the first circulant matrix plus error patterns appearing inside $(6,0)$ trapping sets. Similarly, we can define a new decoder 
$$D_{10}=(\{\Psi_{10}^{\mathcal{V}_1}, \Psi_{10}^{\mathcal{V}_2}\}, \\(0,1,0,0,0,1,1,0,1,0)),$$
where $\Psi_{10}^{\mathcal{V}_1}$ is described by Table~\ref{tab:f11} and $\Psi_{10}^{\mathcal{V}_2}$ is described by Table~\ref{tab:f1}.
Hence, the sets $\{D_1,D_9\}$ and $\{D_{10},D_1\}$ can guarantee the correction of all classical weight-$3$ errors and errors inside symmetric stabilizers. Following the same idea, we can build sets of decoders based on Table~\ref{tab:div} (except $D_1$), where we will complement each of the decoders with two decoders with the same $f$ function and two differing $\Psi$ functions same as $D_9$ and $D_{10}$. Eventually, we will form a set of $24$ decoders that can correct error patterns up to weight-$5$ that appear inside classical trapping sets and error patterns inside quantum trapping sets. As we will observe in the following section, this set of $24$ decoders will help improve both the waterfall and error-floor performance over the previous set of decoders for three different degree-$3$ QLDPC codes. Following the same idea,  one can assign a set of decoders that can simultaneously correct error patterns in each one of the $16$ classical trapping sets of the code.

\section{Numerical results}
\label{sec:sim}
We simulate the decoding performance of the proposed set of decoders given in Section~\ref{sec:dec}. Unless noted otherwise, the maximum number of iterations for any decoder depicted in the upcoming plots is set to $50$. Note that we choose to implement MS decoding because it achieves better performance than BP when a normalizing parameter is applied. 
Also, every decoder shown in the plots follows a parallel-scheduling approach, except noted otherwise (for example, layered decoding).

\begin{figure}[htb]
        \centering
        \includegraphics[width=.4\textwidth]{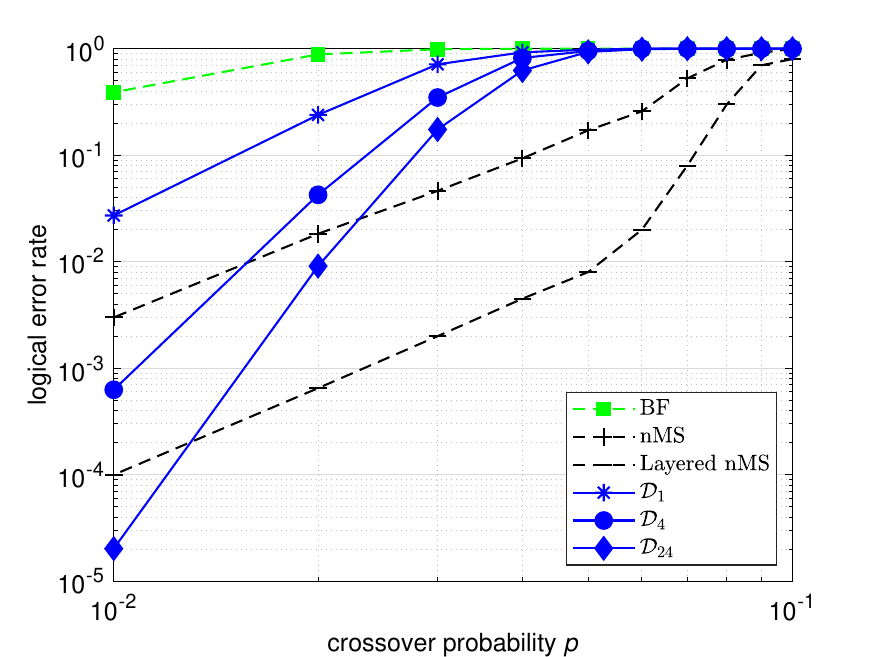}
        \caption{Performance of collective bit flipping-based decoding for the $B1$ code. $\mathcal{D}_4$ surpasses nMS decoding while $\mathcal{D}_{24}$ also surpasses layered nMS decoding in the error-floor region.}
        \label{fig:B1tbfa4}
    \end{figure}

Fig.~\ref{fig:B1tbfa4} illustrates the decoding performance of BF, normalized min-sum (nMS), layered nMS, and three sets of TBF decoders. The performance plot for layered nMS has been replicated from~\cite{layeredQuantum}. We note that the difference between the nMS plot in our implementation and the one in~\cite{layeredQuantum} is based on the fact that in the latter, the authors pick the normalization factor uniformly at random in
$\{0.875,0.9275\}$ for every iteration, whereas we pick it to be equal to $0.875$. The first set of TBF decoders is only comprised of $\mathcal{D}_1=D_1$ which can correct all error patterns up to weight-$3$ that appear inside the $(49,49)$ and $(63,63)$ trapping sets. The second set of TBF decoders is comprised of $\mathcal{D}_4=\{D_1,D_2,D_3,D_9\}$ which are given in Table~\ref{tab:div} and can correct every weight-$4$ error pattern applied in either the $(49,49)$ trapping set or $(63,63)$ trapping set and every error pattern appearing in the $(6,0)$ trapping set independently.
Finally, the third set of TBF decoders $\mathcal{D}_{24}$, is the one described in Section~\ref{sec:decb}.
The poor performance of BF  is attributed to its inability to correct error patterns of weight as low as two (see also Section~\ref{sec:dec}). 
On the other hand, we observe that $\mathcal{D}_4$ outperforms both BF and nMS, while $\mathcal{D}_{24}$ also outperforms the layered nMS, which demonstrates an error-floor behavior. 

         \begin{figure}[htb]
        \centering
        \includegraphics[width=.4\textwidth]{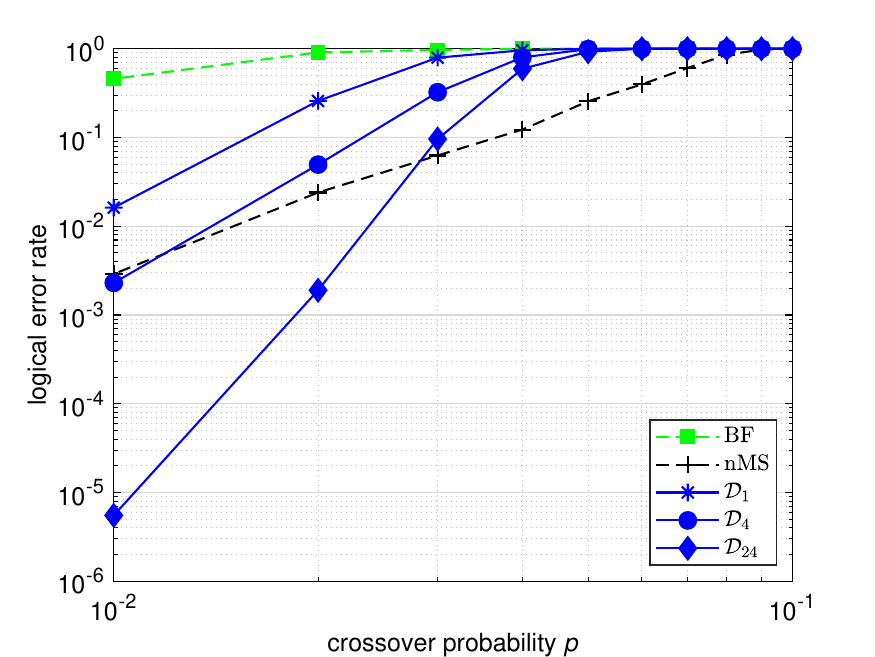}
        \caption{Performance of collective bit flipping-based decoding for the $B3$ code. $\mathcal{D}_{24}$ surpasses nMS decoding. The set of proposed decoders provides better performance than that of the $B1$ code (for which it was originally designed) due to the sparsity that the $B3$ code offers.}
        \label{fig:B124}
    \end{figure}
      \begin{figure}[htb]
        \centering \includegraphics[width=.4\textwidth]{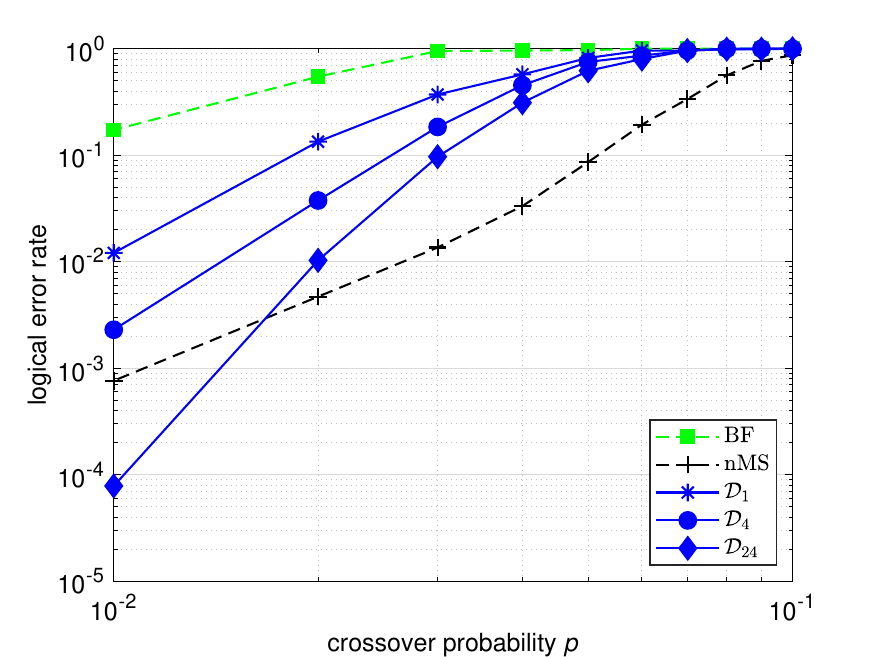}
        \caption{Performance of collective bit flipping-based decoding for the $\llbracket 288, 12, 18 \rrbracket$ BB code. The ensemble of $24$ decoders outperforms the nMS algorithm.}
        \label{fig:BB}
    \end{figure}
         \begin{figure}[htb]
        \centering
    \includegraphics[width=.4\textwidth]{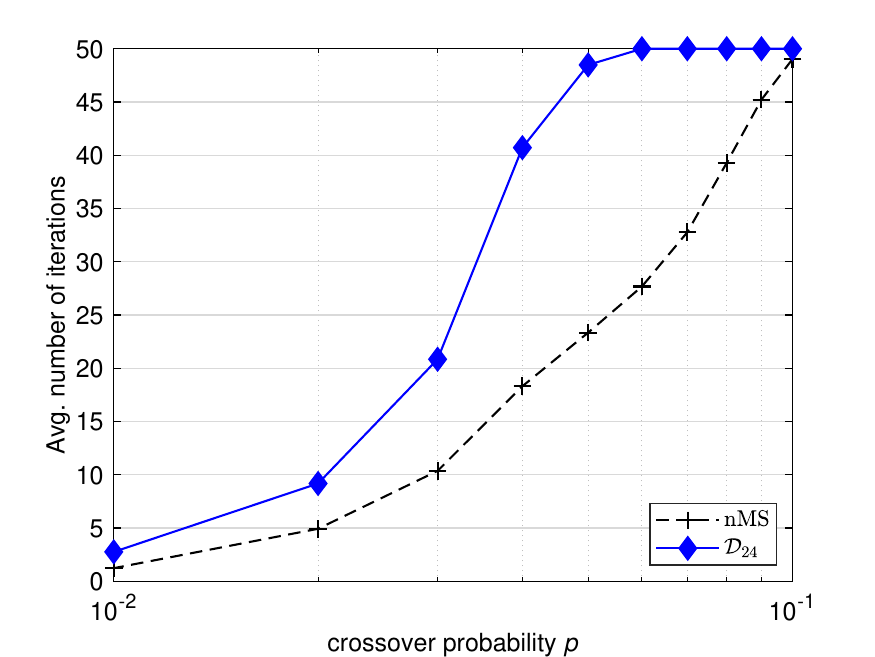}
        \caption{Average number of iterations required by nMS and $\mathcal{D}_{24}$ for decoding the $B3$ code.}
        \label{fig:avgIters}
    \end{figure}
    
For the $B3$ code (Fig.~\ref{fig:B124}), we compare the performance of the ensembles of decoders against the nMS algorithm (using the same normalization factor). We note here that $B3$ code also contains the $(6,0)$ trapping sets, for which we can decode every error pattern that appears inside it (for more details about the $A$ and $B$ matrices that describe $B3$, refer to~\cite{osd}). BF exhibits poor decoding performance, while our proposed set of four decoders ($\mathcal{D}_4$) can almost match the performance of the nMS in the error-floor region. Interestingly, the performance gain of the ensemble of $24$ decoders is higher for the $B3$ code when compared to the $B1$ code. This happens because the one circulant matrix of $B3$ is also composed of juxtaposed $6$-cycles, similar to the $B1$ code, but on the other hand, the other circulant matrix has a girth equal to $10$, which we conjecture that further enhances the performance of our scheme due to its sparsity.

In addition, we choose to use our proposed ensemble of decoders for a family of $d_v$-$3$ QLDPC codes that were recently proposed in~\cite{bravyi2024high}, namely the Bivariate Bicycle (BB) LDPC codes. BB codes are the first example of high-rate, large-distance QLDPC codes achieving the pseudo-threshold close to $1 \%$ under the circuit-based noise model. These codes share a similar structure to the family of GHP that we studied, so we expect our proposed set of decoders to improve their performance as well. We note that the authors provide decoding results by only considering BP-OSD under a circuit-based noise model. In Fig.~\ref{fig:BB} we compare the performance of nMS against that of $\mathcal{D}_1$, $\mathcal{D}_4$, and $\mathcal{D}_{24}$. For the $\llbracket 288, 12, 18 \rrbracket$ code, we notice that collective decoding still outperforms nMS, but the gain over the error correction performance is less than the one observed for the $B1$ and $B3$ codes. This happens because the $\llbracket 288, 12, 18 \rrbracket$ code does not contain the $(63,63)$ and $(49,49)$ trapping sets. Nevertheless, since BB codes still contain the $(6,0)$ trapping sets and their classical trapping sets are formed by juxtaposing $6$-cycles, $\mathcal{D}_{24}$ performs about $10$ times better than nMS for a crossover probability equal to $0.01$. We also note that no error-floor behavior is observed for any of the three sets of decoders.

In general, the way $6$-cycles are juxtaposed is important. For example, to attain the same error correction capability for the $(49,49)$ and $(63,63)$ trapping sets independently, the former required fewer TBF decoders. This behavior is attributed to the $(49,49)$ trapping set having a sparser structure than the $(63,63)$ trapping set, as it contains longer chains of interconnected $6$-cycles. We conjecture that correcting error patterns that appear inside structures formed by interconnected $6$-cycles is essential for decoding $d_v$-$3$, $d_c$-$6$ QLDPC codes, and we showed that our proposed ensembles can provide good error-floor performance for a wide range of such codes. A possible future research direction would be to define decoders that solve many kinds of juxtapositions of $6$-cycles and then deploy them together to decode a wide variety of codes.

Finally, Fig.~\ref{fig:avgIters} demonstrates the average number of decoding iterations required by $\mathcal{D}_{24}$ \emph{vs.} the nMS decoder for the $B3$ code (corresponding to Fig.~\ref{fig:B124}). We observe that for a crossover probability of $0.01$, $\mathcal{D}_{24}$ attains about $1000$ times better performance than nMS by only requiring about $2.5$ iterations, while nMS requires $1.2$ iterations. We note that a TBF-based decoding iteration is expected to require less latency than a nMS iteration; therefore, the collective decoding approach provides a good performance/latency trade-off.
  
\section{Conclusions and future directions}
We presented a decoding method that applies to families of QLDPC codes with specific characteristics and can improve their error-floor performance under low-latency requirements. Our proposed decoder is a product of the trapping set analysis of $d_v$-$3$ QLDPC codes and comprises multiple TBF decoders that run in parallel and can collectively correct error patterns appearing inside harmful configurations. Performance results demonstrated that our approach significantly outperforms MS, mainly in the error-floor region, and can do so in fewer iterations (on average, two iterations for a crossover probability of $0.01$), thus addressing the stringent latency requirements of quantum decoding. Future work will be focused on further exploiting the structure of QLDPC codes, i.e., by assigning different decoding rules to variable nodes regarding the trapping set they belong to, which will, in turn, enable the decoding of higher-weight error patterns. Also, we aim to extend the ideas of diversity to stronger decoders and combine low-latency decoding and good waterfall performance.
\section*{Acknowledgments}
This work is supported by the NSF under grants CIF-2106189, CCF-2100013, ECCS/CCSS-2027844, ECCS/CCSS-2052751, and in part by the CoQREATE program under grant ERC-1941583, and Jet Propulsion Laboratory, California Institute of Technology, under a contract with the National Aeronautics and Space Administration and funded through JPL’s Strategic University Research Partnerships (SURP) program. Bane Vasi\'{c} has disclosed an outside interest in his startup company, Codelucida to The University of Arizona. Conflicts of interest resulting from this interest are being managed by The University of Arizona in accordance with its policies.  

The authors are with the Department of Electrical and Computer Engineering, The University of Arizona, Tucson, AZ 85721, USA (e-mail: dchytas@arizona.edu; nithin@arizona.edu; vasic@ece.arizona.edu).

\IEEEtriggeratref{25} 
    \bibliographystyle{IEEEtran}
\bibliography{references}
\end{document}